\documentclass[10pt,twoside]{article}
\usepackage{Latex-document}

\usepackage{amssymb}
\usepackage{amsfonts}
\usepackage{amsmath}

\usepackage{amsthm}

\usepackage{float}

\renewcommand{\leq}{\leqslant}

\theoremstyle{plain}
\newtheorem{theorem}{Theorem}
\newtheorem{lemma}[theorem]{Lemma}
\newtheorem{corollary}[theorem]{Corollary}

\theoremstyle{definition}

\floatstyle{boxed}
\newfloat{algorithm}{thpb}{loa}
\floatname{algorithm}{Algorithm}

\title{\bf Hardness as  Randomness: \vskip -2mm A Survey of Universal Derandomization\thanks{Research
supported by NSF Award CCR-0098197 and USA-Israel BSF Grant
97-00188.}\vskip 6mm}

\author{Russell Impagliazzo\thanks{Department of Computer Science, University of California, San Diego,
La Jolla, CA 92093-0114, USA. E-mail: russell@cs.ucsd.edu}}

\date{\vspace{-8mm}}

\begin{document}

\maketitle

\thispagestyle{first} \setcounter{page}{659}

\begin{abstract}

\vskip 3mm

We survey recent developments in the study of probabilistic complexity classes.  While the evidence seems to
support the conjecture that probabilism can be deterministically simulated with relatively low overhead, i.e.,
that $P=BPP$, it also indicates that this may be a difficult question to resolve.  In fact, proving that
probalistic algorithms have non-trivial deterministic simulations is basically equivalent to proving circuit lower
bounds, either in the algebraic or Boolean models.

\vskip 4.5mm


\end{abstract}

\vskip 12mm

\section{Introduction}

\vskip-5mm \hspace{5mm}

The use of random choices in algorithms has been a suprisingly
productive idea.   Many problems that have no known efficient
deterministic algorithms have fast randomized algorithms, such as
primality and polynomial identity testing. But to what extent is
this seeming power of randomness real? Randomization is without
doubt a powerful algorithm design tool, but does it dramatically
change the notion of efficient computation?

To formalize this question, consider $BPP$, the class of problems
solvable by bounded error probabilistic polynomial time
algorithms. It is possible that $P=BPP$, i.e., randomness never
solves new problems. However, it is also possible that $BPP=EXP$,
i.e., randomness is a nearly omnipotent algorithmic tool.

Unlike for $P vs. NP$, there is no consensus intuition concerning
the status of $BPP$. However, recent research gives strong
indications that adding randomness does not in fact change what is
solvable in polynomial-time, i.e., that $P=BPP$. Surprisingly, the
problem is strongly connected to circuit complexity, the question
of how many operations are required to compute a function.

{\em A priori}, possibilities concerning the power of randomized
algorithms include:
\begin{enumerate}
\item  Randomization always helps for intractable problems,
i.e., $EXP=BPP$.
\item
The extent to which randomization helps is problem-specific. It
can reduce complexity by any amount from not at all to
exponentially.
\item
True randomness is never needed, and random choices can always be
simulated deterministically, i.e., $P=BPP$.
\end{enumerate}

Either of the last two possibilities seem plausible, but most
consider the first wildly implausible. However, while a strong
version of the middle possibility has been ruled out, the
implausible first one is still open. Recent results indicate both
that the last, $P=BPP$, is both very likely to be the case and
very difficult to prove.

More precisely:
\begin{enumerate}
\item  Either no problem in $E$ has strictly exponential
circuit complexity or $P=BPP$. This seems to be strong evidence
that, in fact, $P=BPP$, since otherwise circuits can always
shortcut computation time for hard problems.
\item Either $BPP=EXP$, or any problem in $BPP$ has
a deterministic sub-exponential time algorithm that works on
almost all instances. In other words, either randomness solves
every hard problem, or it does not help exponentially, except on
rare instances. This rules out strong problem-dependence, since if
randomization helps exponentially for many instances of {\em some
problem}, we can conclude that it helps exponentially for {\em all
intractible problems}.
\item If $BPP=P$, then either the permanent problem
requires super-polynomial algebraic circuits or there is a problem
in $NEXP$ that has no polynomial-size Boolean circuit. That is,
proving the last possibility requires one to prove a new circuit
lower bound, and so is likely to be difficult.
\end{enumerate}

The above are joint work with Kabanets and Wigderson, and use
results from many others.

All of these results use the hardness-vs-randomness paradigm
introduced by Yao \cite{Yao}: Use a hard computational problem to
define a small set of ``pseudo-random'' strings, that no limited
adversary can distinguish from random.  Use these
``pseudo-random'' strings to replace the random choices in a
probabilistic algorithm. The algorithm will not have enough time
to distinguish the pseudo-random sequences from truly random ones,
and so will behave the same as it would given random sequences.

In this paper, we give a summary of recent results relating
hardness and randomness.  We explain how the area drew on and
contributed to coding theory, combinatorics, and structural
complexity theory. We will use a very informal style.  Our main
objective is to give a sense of the ideas in the area, not to give
precise statements of results. Due to space and time limitations,
we will be omitting a vast amount of material. For a more complete
survey, please see \cite{Kab02-beatcs}.

\section{Models of computation and complexity classes}

\vskip-5mm \hspace{5mm}

The $P$ vs. $BPP$ question arises in the broader context of the
robustness of models of computation. The famous Church-Turing
Thesis states that the formal notion of recursive function
captures the conceptual notion of computation.   While this is not
in itself a mathematical conjecture, it has been supported by
theorems proving that various ways of formalizing
``computability'', e.g., Turing Machines and the lambda calculus,
are in fact equivalent.

When one considers complexity as well as computability, it is
natural to ask if a model also captures the notion of computation
time.  While it became apparant that exact computation time was
model-dependent, simulations between models almost always
preserved time  up to a polynomial. The time-restricted
Church-Turing thesis is that any two reasonable models of
computation should agree on time up to polynomials; equivalently,
that the class of problems decideable in polynomial time be the
same for both models.  For many natural models, this is indeed the
case, e.g. RAM computation, one-tape Turing machines, multi-tape
Turing machines, and Cobham's axioms all define the same class $P$
of poly-time decideable problems.

Probabilistic algorithms for a long time were the main challenge
to this time-restricted Church-Turing thesis. If one accepts the
notion that making a fair coin flip is a legitimate, finitely
realizable computation step, then our model of poly-time
computation seems to change. For example, primality testing
\cite{SS,Rab} and polynomial identity testing \cite{Sch80, Zip79}
are now polynomial-time, whereas we do not know any deterministic
polynomial-time algorithms.  The $P$ vs. $BPP$ question seeks to
formalize the question of whether this probabilistic model is
actually a counter-example, or whether there is some way to
simulate randomness deterministically.

As a philisophical question, the Church-Turing Thesis has some
ambiguities. We can distinguish at least two variants: a
conceptual thesis that the standard model captures the conceptual
notion of computation and computation time, and a physical thesis
that the model characterizes the capabilities of
physically-implementable computation devices.  In the latter
interpretation, quantum physics is inherrently probabilistic, so
probabilistic machines seem more realistic than deterministic ones
as such a characterization.

Recently, researchers have been taking this one step further by
studying models for quantum computation. Quantum computation is
probably an even more serious challenge to the time-limited
Church-Turing thesis than probabilistic computation.  This lies
beyond the scope of the current paper, except to say that we do
not believe that any analagous notion of pseudo-randomness can be
used to deterministically simulate quantum algorithms. Quantum
computation is intrinsically probabilistic; however, much of its
power seems to come from interference between various possible
outcomes, which would be destroyed in such a simulation.

\subsection{Complexity classes}

\vskip-5mm \hspace{5mm}

We assume familiarity with the standard deterministic and
non-deterministic computation models (see \cite{Pap94} for
background.)  To clarify notation, $P\ =\ $$DTIME(n^{O(1)})$ is
the class of decision problems solvable in deterministic
polynomial time, $E =DTIME (2^{O(n)})$ is the class of such
problems decideable in time exponential in the input length, and
$EXP=DTIME(2^{n^{O(1)}})$ is the class of problems solvable in
time exponential in a polynomial of the input length. $NP,NE,$ and
$NEXP$ are the analogs for non-deterministic time.  If $C_1$ and
$C_2$ are complexity classes, we use $Co-C_1$ to denote the class
of complements to problems in $C_1$, and $C_1^{C_2}$ to represent
the problems solvable by a machine of the same type as normally
accept $C_1$, but which is also allowed to make oracle queries to
a procedure for a fixed language in $C_2$. ( This is not a precise
definition, and to make it precise, we would usually have to refer
to the definition of $C_1$.  However, it is also usually clear
from context how to do this.) The {\em polynomial hierarchy} $PH$
is the union of $NP,\Sigma^P_2 = NP^NP,
\Sigma^P_3=NP^{\Sigma^P_2},...$.

A {\em probabilistic algorithm} running in $t(|x|)$ time is an
algorithm $A$ that uses, in addition to its input $x$, a randomly
chosen string $r \in \{0,1\}^{t(n)}$. Thus, $A(x)$ is a
probability distribution on outputs $A(x,r)$ as we vary over all
strings $r$. We say that $A$ {\em recognizes} a language $L$ if
for every $x \in L$, $Prob[A(x,r)=1] > 2/3$ and every $x \not\in
L$, $Prob[A{x,r}=1] < 1/3$, where probabilities are over the
random tape $r$. $BPP$ is the class of languages recognized by
polynomial-time probabilistic algorithms.

The gap between probabilities for acceptance and rejection ensures
that there is a statistically significant difference between
accepting and rejecting distributions.  Setting the gap at 1/3 is
arbitrary; it could be anything larger than inverse polynomial,
and smaller than $1-$ an inverse exponential, without changing the
class $BPP$.  However, it does mean that there are probabilistic
algorithms, perhaps even useful ones, that do not accept any
language at all. Probabilistic heuristics might clearly accept on
some inputs, clearly reject on others, but be undecided sometimes.

To handle this case, we can introduce a stronger notion of
simulating probabilistic algorithms than solving problems in
$BPP$.  Let $A$ be any probabilistic algorithm.  We say that a
deterministic algorithm $B$ solves the {\em promise problem} for
$A$ if, $B(x)=1$ whenever $Prob[A(x,r)=1]> 2/3$ and $B(x)=0$
whenever $Prob[A(x,r)=1]< 1/3$.  Note that, unlike for $BPP$
algorithms, there may be inputs on which $A$ is basically
undecided; for these $B$ can output either $0$ or $1$. We call the
class of promise problems for probabilistic polynomial time
machines $Promise-BPP$.  Showing that $Promise-BPP \subseteq P$ is
at least as strong and seems stronger than showing $BPP=P$. (See
\cite{For01,KRC00-eccc} for a discussion.)

As happens frequently in complexity, the negation of a good
definition for ``easy'' is not a good definition for ``hard''.
While $EXP=BPP$ is a good formalization of ``Randomness always
helps'', $BPP = P$ is less convincing as a translation of
``Randomness never helps''; $Promise-BPP \subseteq P$ is a much
more robust statement along these lines.

$\#P$ is the class of counting problems for polynomail-time
verifiable predicates. i.e., For each poly-time predicate $B(x,y)$
and polynomial $p$, the associated counting problem is: given
input $x$, how many $y$ with $|y|=p(|x|)$ satisfy $B(x,y)=1$?
Valiant showed that computing the permanent of a matrix is
$\#P$-complete \cite{Val79-stoc}, and Toda showed that $PH
\subseteq P^{\#P}$ \cite{Toda}.

A class that frequently arises in proofs is $MA$, which consists
of languages with probabilistically verifiable proofs of
membership. Formally, a language $L$ is in $MA$ if there is a
predicate $B(x,y,r)$ in $P$ and a polynomial $p$ so that, if $x
\in L$, $\exists y  |y|=p(|x|)$ so that $Prob_{r \in_U
\{0,1\}^{p(|x|}} [B(x,y,r)=1] >2/3$ and if $x \not\in L$, $\forall
y, |y|=p(|x|)$, $Prob_{r \in_U \{0,1\}^{p(|x|}} [B(x,y,r)=1] <
1/3$. Although $MA$ combines non-determinism and probabilism,
there is no direct connection known between derandomizing $BPP$
and derandomizing $MA$.  This is because if $x \in L$, there still
may be some poorly chosen witnesses $y$ which are convincing to
$B$ about 1/2 the time. However, derandomizing $Promise-BPP$ also
derandomizes $MA$, because we don't need a strict guarantee.
\begin{lemma}
Let $T(n)$ be a class of time-computable functions closed under
composition with polynomials. If $Promise-BPP \subseteq
NTIME[T(n)]$ then $MA \subseteq NTIME[T(n)]$ .
\end{lemma}

\subsection{Boolean and algebraic circuits}

\vskip-5mm \hspace{5mm}

The circuit complexity of a finite function measures the number of
primitive operations needed to compute the function. Starting with
the input variables, a circuit computes a set of intermediate
values in some order.  The next intermediate value in the sequence
must be computed as a primitive  operation of the inputs and
previous intermediate values.  One or more of the values are
labelled as outputs; for one output circuits this is without loss
of generality the last value to be computed. The size of a circuit
is the number of values computed, and the circuit complexity of a
function $f$, $Size(f)$, is the smallest size of a circuit
computing $f$.

Circuit models differ in the type of inputs and the primitive
operations.  Boolean circuits have Boolean inputs and the Boolean
functions on  1 or 2 inputs as their primitive operations.
Algebraic circuits have inputs taking values from a field $G$ and
whose primitive operations are addition in $G$, multiplication in
$G$, and the constants  $1$ and $-1$.  Algebraic circuits can only
compute polynomials. Let $f_n$ represent the function $f$
restricted to inputs of size $n$ We use the notation $P/poly$ to
represent the class of functions $f$ so that the Boolean circuit
complexity of $f_n$ is bounded by a polynomial in $n$; we use the
notation $AlgP/poly$ for the analagous class for algebraic
circuits over the integers.

Circuits are {\em non-uniform} in that there is no a priori
connection between the circuits used to compute the same function
on different input sizes.  Thus, it is as if a new algorithm can
be chosen for each fixed input size. While circuits are often
viewed as a combinatorial tool to prove lower bounds on
computation time, circuit complexity is also interesting in
itself, because it gives a concrete and non-asymptotic measure of
computational difficulty.

\section{Converting hardness to pseudorandomness}

\vskip-5mm \hspace{5mm}

To derandomize an algorithm $A$, we need to, given $x$,  estimate
the fraction of strings $r$ that cause probabilistic algorithm
$A(x,r)$ to output 1. If $A$ runs in $t(|x|)$ steps, we can
construct an approximately $t(|x|)$ size circuit $C$ which on
input $r$ simulates $A(x,r)$.  So the problem reduces to: given a
size $t$ circuit $C(r)$, estimate the fraction of inputs on which
it accepts. Note that solving this circuit-estimation problem
allows us to derandomize $Promise-BPP$ as well as $BPP$.

We could solve this by searching over all $2^t$ $t$-bit strings,
but we'd like to be more efficient. Instead, we'll search over a
specially chosen small {\em sample set} $S=\{r_1,...r_s\}$ of such
strings. The average value over $r_i \in S$ of $C(r_i)$
approximate the average over all $r$'s  for any small circuit $C$.
This is basically the same as saying that the task of
distinguishing between a random string and a member of $S$ is so
computationally difficult that it lies beyond the abilities of
size $t$ circuits.  We call such a sample set {\em pseudo-random}.
Pseudo-random sample sets are usually described as the range of a
function called a {\em pseudo-random generator}.  This made sense
for the original constructions, which had cryptographic
motivations, and where it was important that $S$ could be sampled
from very quickly \cite{BM,Yao}. However, we think the term
pseudo-random generator for hardness vs. randomness is merely
vestigial, and in fact has misleading connotations, so we will use
the term {\em pseudo-random sample set}.

We want to show the existence of a function with small

Since we want distinguishing members of $S$ to be hard for all
small circuits, we need to start with a problem $f$ of high
circuit complexity, say $Size(f) \ge t^{c}$ for some constant $c >
0$. We assume that we have or compute the entire truth table for
$f$.

For the direct applications, we'll obtain $f$ as follows.  Start
with some function $F \in E$ defined on all input sizes, where
$F_{\eta}$ is has circuit size at least $H(\eta)$ for a
super-polynomial function $H$.  Pick $\eta$ so that $H(\eta) \ge
t^c$ and let $f=F_{\eta}$. Note that $t^{o(1)} \ge \eta > \log t$.
Since $F \in E$, we can construct the truth-table for $f$ in time
exponential in $\eta$, which means polynomial time in the size of
the truth-table, $n=2^{\eta}$.

Other applications, in later sections, will require us to be able
to use any hard function, not necessarily obtained from a fixed
function in $E$.

We then construct from $f$ the pseudo-random sample set $S_f
\subseteq \{0,1\}^t$.  Given the truth table of $f$, we list the
members of $S_f$ in as small a deterministic time as possible.  It
will almost always be possible to do so in time polynomial in the
number of such elements, so our main concern will be minimizing
the size of $S_f$.  We then need to show that no $t$ gate circuit
can distinguish between members of $S_f$ and truly random
sequences. We almost always can do so in a very strong sense:
given a test $T$ that distinguishes $S_f$ from the uniform
distribution, we can produce a size $t^{c-1}$ size circuit using
$T$ as an oracle, $C^T$,  computing $f$. If such a test were
computable in size $t$, we could then replace the oracle with such
a circuit, obtaining a circuit of size $t^c$ computing $f$, a
contradiction.

The simulation is:  Choose $\eta$.  Construct the truth table of
$f=F_{\eta}$. Construct $S_f$.  Run $A(x,r_i)$ for each $r_i \in
S_f$. Return the majority answer.  In almost all constructions,
the dominating term in the simulation's time is the size of $S_f$.
In the most efficient constructions, making the strongest hardness
assumption, $H(\eta) \in 2^{\Omega(\eta)}$, \cite{IW97, STV01}
obtain constructions with $|S_f| =n^{O(1)}=t^{O(1)}$. This gives
us the following theorem:
\begin{theorem}
If there is an $F \in E$ with $Size(F_\eta) \in 2^{\Omega(\eta)}$
then $P=BPP$.
\end{theorem}

\cite{Uma02} gives an optimally efficient construction for any
hardness, not just exponential hardness.

\subsection{The standard steps}

\vskip-5mm \hspace{5mm}

The canonical outline for constructing the pseudo-random sample
set was first put together in \cite{BFNW};  however, each of their
three steps was at least implicit in earlier papers.  Later
constructions either improve one of the steps, combine steps, or
apply the whole argument recursively.  However, a conceptual
break-through that changed the way researchers looked at these
steps is due to \cite{Tre01} and will be explored in more detail
in the next section.
\begin{enumerate}
\item Extension and random-self-reduction.  Construct
from $f$ a function $\hat{f}$ so that, if $\hat{f}$ has a circuit
that computes its value correctly on {\em almost all} inputs, then
$f$ has a small circuit that is correct on {\em all} inputs.
\\
This is usually done by viewing $f$ as a multi-linear or
low-degree polynomial over some field of moderate characteristic
(poly in $\eta$). Then that polynomial can be extrapolated to
define it at non-Boolean inputs, giving the extension $\hat{f}$.
If we have a circuit that is almost always correct, we can produce
a probabilistic circuit that is always correct as follows.  To
evaluate $\hat{f}$ at $v$, pick a point $w$ at random, and
evaluate the almost always correct circuit at random points on the
line $l=v+x*w$. Since any  point is on exactly one line with $v$,
these points are uniform, and chances are the circuit is correct
on these points.  $\hat{f}$ restricted to $l$ can be viewed as a
low-degree polynomial in the single variable $x$.  Thus, we can
interpolate this polynomial, and use its value at $x=0$ to give us
the value $\hat{f}(v)$.  (\cite{BF90} is the first paper we know
with this construction.)
\\
The key parameter  that influences efficiency for this stage is
$\hat{\eta}$, since the size of the truth-table for $\hat{f}$ is
$\hat{n}=2^{\hat{\eta}}$. Ideally, $\hat{\eta} \in O(\eta)$, so
that $\hat{n} \in n^{O(1)}$, and we can construct $\hat{f}$ in
polynomial-time.
\item Hardness Amplification:  From $\hat{f}$, construct
a function $\overline{f}$ on inputs of size $\overline{\eta}$ so
that, from a circuit that can predict $\overline{f}$ with an
$\epsilon$ advantage over guessing, we can construct a circuit
that computes $\hat{f}$ on almost all inputs.
\\
The prototypical example of a hardness amplification construction
is the exclusive-or lemma \cite{Yao, Le1}. Here
$\overline{f(y_1\circ y_2...\circ y_k)} = \hat{f} (y_1) \oplus
\hat{f} (y_2) ... \oplus \hat{f} (y_k)$. Efficiency for this stage
is mostly minimizing $\hat{\eta}$. The $\oplus$ construction above
is not particularly efficient, so much work went into more
efficient amplification.

\item Finding quasi-independent sequences of inputs.
Now we have a function whose outputs are almost as good as random
bits at fooling a size-limited guesser. However, we need many
output bits that look mutually random.  In this step, a small sets
of input vectors $V$ is constructed so that for $(v_1,...v_t)
\in_U V$, guessing $\overline{f}$ on $v_i$ is hard and in some
sense independent of the guess for $v_j$.
\\
Then the sample set will be defined as: $S=
\{(\overline{f}(v_1),...\overline{f}(v_t)) | (v_1,...v_t) \in V\}$
\\
The classical construction for this step is from \cite{NW}.  This
construction starts with a {\em design}, a family  of subsets
$D_1,..D_t \subseteq [1, .. \mu] , |D_i|= \overline{\eta}$, and
$|D_i \cap D_j| \leq \Delta$ for  $i \neq j$.  Then for each $w
\in \{0,1\}^{\mu}$ we construct $v_1,...v_t$, where $v_i$ is the
bits of $w$ in $D_i$, listed in order.  Intuitively, each $v_i$ is
``almost independent'' of the other $v_j$, because of the small
intersections. More precisely, if a test predicts  $\hat{f}(v_i)$
from the other $v_j$, we can restrict the parts of $w$ outside
$D_i$.  Then each restricted $v_j$ takes on at most $2^{\Delta}$
values, but we haven't restricted $v_i$ at all. We can construct a
circuit that knows these values of $\hat{f}$  and uses them in the
predictor.
\\
The size of $S_f$ is $2^{\mu}$, so for efficiency we wish to
minimize $\mu$. However, our new predicting circuit has size
$2^{\Delta} poly(t)$, so we need $\Delta \in O(\log t)$.  Such
designs are possible if and only if $\mu \in \Omega(
\overline{\eta}^2/ \Delta)$. Thus, the construction will be
poly-time if we can have $\overline{\eta} = O(\eta)=O(\log t)$.
\end{enumerate}

\section{Extractors, Graphs, and Hardness vs. Randomness}

\vskip-5mm \hspace{5mm}

As mentioned before, \cite{Tre01} changed our persective on
hardness vs. randomness. We mentioned earlier that it was
plausible that nature had truly probabilistic events.  But is it
plausible that we can physically construct a perfect fair coin?
Many physical sources of randomness have imperfections and
correlations. From the strong versions of hardness vs. randomness
constructions, we can simulate a randomized algorithm without
making the assumption that perfect random bits are available.  Say
we are simulating a randomized algorithm using $t$ perfect random
bits. (We don't need to have a time bound for the algorithm). Let
$T$ be the set of random sequences on which the algorithm accepts.

Assume we have a physical source outputting $n$ bits, but all we
know about it is that no single output occurs more than
$2^{-t^{c+1}}$ of the time, i.e., that it has min-entropy at least
$t^{c+1}$. Treating the output of the source as a function $f$ on
$\eta = \log n$ bits, we construct the sample set $S_f$, and
simulate the algorithm on the sample set.  The min-entropy and a
simple counting argument suffices to conclude that most outputs do
not have small circuits relative to $T$. Therefore, most outputs
of the source have about the right number of neighbors in $T$, and
so our simulation works with high probability.

This connection has been amazingly fruitful, leading to better
constructions of extractors as well as better hardness vs.
randomness results.

This construction is also interesting from the point of view of
quasi-random graphs. Think globally.  Instead of looking at the
sample set construction on a single function $f$, look at it on
all possible functions.

This defines a bipartite graph, where on the right side, we have
all $2^{2^{\eta}}=2^n$ functions on $\eta$ bits, and on the left
side, we have all $t$ bit strings; the edges are between each
function $f$ and the members of the corresponding sample set
$S_f$. Let $T$ be any subset of the left side.  Then we know that
any function $f$ that has many more or fewer than $s |T|/2^t$
neighbors in $T$ has small circuit complexity relative to $T$.  In
particular, there cannot be too many such functions.
Contrapositively, any large set of functions must have about the
right number of neighbors in $T$. Thus, we get a combinatorially
interesting construction of an extremely homogenous bipartite
graph from any hardness vs. randomness result.

\subsection{The steps revisited}

\vskip-5mm \hspace{5mm}

Once we look at the hardness vs. randomness issue from the point
of view of extracting randomness from a flawed source, we can
simplify our thoughts about the various steps.  Any particular
bits, and even most bits, from a flawed random source might be
constant, because outputs might tend to be close in Hamming
distance. This problem suggests its own solution:  Use an error
correcting code first. Then any two outputs are far apart, so most
bit positions will be random.  In fact, in retrospect, what the
first two steps of the standard hardness vs. randomness method are
doing is error-correcting the function. We do not care very much
about rate, unless the rate is not even inverse polynomial.
However, we want to be able to correct even if there is only a
slight correlation between the recieved coded message and the
actual coded message.  It is information-theoretically impossible
to uniquely decode under such heavy noise, but it is sometimes
possible to {\em list decode}, producing a small set of possible
messages.  At the end of the hardness amplification stage, this is
in fact what we have done to the function.

However, there are some twists to standard error- correction that
make the situation unique. Most interestingly, we need decoding
algorithms that are super-fast, in that to compute any particular
bit of the original message can be done in poly-log time, assuming
random access to the bits of the coded message.  This kind of {\em
local decodability} was implicit in \cite{AS97}, and applied to
hardness vs. randomness in \cite{STV01}.

In retrospect, much of the effort in hardness-vs-randomness
constructions has been in making locally list-decodeable
error-correcting codes in an ad hoc manner. \cite{STV01} showed
that even natural ways of encoding can be locally list-decodeable.
However, there might be some value in the ad hoc approaches.  For
example, many of the constructions assume the input has been
weakly error-corrected, and then do a further construction to
increase the amount of noise tolerated. Thus, these constructions
can be viewed as error-correction boosters:  codes where, given a
code word corrupted with noise at a rate of $\gamma$, one can
recover not the original message, but a message of lower relative
noise, i.e. Hamming distance $\delta n$ from the original message,
where $\delta < \gamma$. These might either be known or of
interest to the coding community.

\section{Hardness from derandomization}

\vskip-5mm \hspace{5mm}

Are circuit lower bounds necessary for derandomization? Some
results that suggested they might not be  are \cite{IW98} and
\cite{Kab01-jcss}, where average-case derandomization or
derandomization vs. a deterministic adversary was possible based
on a uniform or no assumption.  However, intuitively, the instance
could code a circuit adversary in some clever way, so worst-case
derandomization based on uniform assumptions seemed difficult.
Recently, we have some formal confirmation of this:  Proving
worst-case derandomization results automatically prove new circuit
lower bounds.

These proofs usually take the contrapositive approach. Assume that
a large complexity class has small circuits. Show that randomized
computation is unexpectedly powerful as a result, so that the
addition of randomness to a class jumps up its power to a higher
level in a time hierarchy.  Then derandomization would cause the
time hierarchy to collapse, contradicting known time hierarchy
theorems.

An example of unexpected power of randomness when functions have
small circuits is the following result from \cite{BFNW}:
\begin{theorem}
If $EXP \subseteq P/poly$, then $EXP=MA$.
\end{theorem}
This didn't lead directly to any hardness from derandomization,
because $MA$ is the probabilistic analog of $NP$, not of $P$.
However, combining this result with Kabanet's easy witness idea
(\cite{Kab01-jcss}), \cite{IKW01} managed to extend it to $NEXP$.
\begin{theorem}
If $NEXP \subseteq P/poly$, then $NEXP=MA$.
\end{theorem}

Since as we observed earlier, derandomizing $Promise-BPP$
collapses $MA$ with $NP$, it does follow that full derandomization
is not possible without proving a circuit lower bound for $NEXP$.
\begin{corollary}
If $Promise-BPP \subseteq NE$, then $NEXP \not\subseteq P/poly$.
\end{corollary}

A very recent unpublished observation of Kabanets and Impagliazzo
is that the problem of, given an arithmetic circuit $C$ on $n^2$
inputs, does it compute the permanent function. is in $BPP$.  This
is because one can set inputs to constants to set circuits that
should compute the permanent on smaller matrices, and then use the
Schwartz-Zippel test (\cite{Sch80, Zip79}) to test that each
function computes the expansion by minors of the previous one.
Then assume $Perm \in AlgP / poly$.  It follows that $PH \subseteq
P^{Perm} \subseteq NP^{BPP}$, because one could
non-deterministically guess the algebraic circuit for Perm and
then verify one's guess in $BPP$. Thus, if $BPP =P$ (or even $BPP
\subseteq NE$) and $Perm \in AlgP/poly$, then $PH \subseteq NE$.
If in addition, $NE \subseteq P/poly$, we would have $Co-NEXP=NEXP
=MA \subseteq PH \subseteq NE$, a contradiction to the
non-deterministic time hierarchy theorems. Thus, if $BPP \subseteq
NE$, either $Perm \not\in AlgP/poly$ or $NE \not\subseteq P/poly$.
In either case, we would obtain a new circuit lower bound.

\section{Conclusions}

\vskip-5mm \hspace{5mm}

This is an area with a lot of ``good news/bad news'' results. While the latest results seem pessimistic about
finally resolving the $P$ vs. $BPP$ question, the final verdict is still out. Perhaps $NE$ is high enough in
complexity that proving a circuit lower bound there would not require a major breakthrough, only persistance.
Perhaps derandomization will lead to lower bounds, not the other way around. In any case, derandomization seems to
be a nexus of interesting connections between complexity and combinatorics.

\newcommand{\etalchar}[1]{$^{#1}$}

\label{lastpage}

\end{document}